\documentclass[pra,aps]{revtex4}
\usepackage{graphicx}

\begin{document}
\title{Casimir Effect for Massive Scalar Field}
\author{S. Mobassem}
\address{Department of Physics, University of Kerman, Kerman, Iran}
\date{\today}

\begin{abstract}
The energy momentum tensor is used to introduce the Casimir force
of the massive scalar field acting on a nonpenetrating surface.
This expression can be used to evaluate the vacuum force by
employing the appropriate field operators. To simplify our
formalism we also relate the vacuum force expression to the
imaginary part of the Green function via the fluctuation
dissipation theorem and Kubo's formula. This allows one to
evaluate the vacuum force without resorting to the process of
field quantization. These two approaches are used to calculate
the attractive force between two nonpenetrating plates. Special
attention is paid to the generalization of the formalism to $D+1$
space-time dimensions.
\end{abstract}
\maketitle
\section{Introduction}

Historically the Casimir effect is known as the small force
acting between two parallel uncharged perfect conducting plates.
The attractive force per unit area is \cite{ref1}

\begin{equation}
\label{eq1} F=\frac{\pi^2 \hbar c }{240a^4}
\end{equation}
where $\hbar$ is Plank's constant divided by $2\pi$, $c$ is the
velocity of light and $a$ is the separation between the two
conducting plates. Though the force can be thought of as a limiting
case of the Van der Waals interaction when retardation is
included \cite{ref2}, it does not depend on the electric charge.
This encourages one to look for a new interpretation of the force.
It is known that the phenomena stems from the quantum aspect of
the electromagnetic field, since the force tends to zero when we
try to find the classical limit. The force can, therefore, be
considered as the reaction of the vacuum of the quantum field
against the presence of the boundary surface or
surfaces \cite{ref3,ref4,ref5}. In other words, the effect is
simply the stress on the boundary surfaces due to the confinement
of a quantum field in a finite volume of space. The restriction on
the modes of the quantum field gives rise to the force acting on
the macroscopic bodies \cite{ref6}.

Great variety of problems of practical and theoretical interests
are intimately connected to various quantum fields having mass or
massless, quanta of different spin. The later interpretation of
the Casimir force signifies that the effect can not be merely for
the electromagnetic field, and all the other quantum fields must
display the same effect \cite{ref7,ref8,ref9,ref10,ref11}. It is,
in fact, the most important characteristic of the Casimir effect
in all instances that it depends neither on electric charge nor on
any other coupling constants, but it does build on the quantum
nature of the fields \cite{ref12}. The effect can be displayed in
$D+1$ space-time dimensions for a variety of boundary surfaces
and boundary conditions, and manifested on all scales, from the
substructure of quarks \cite{ref13,ref14,ref15,ref16,ref17} to
the large scale structure of the universe \cite{ref18,ref19,ref20,ref21,ref22,ref23,ref24,ref25,ref26}.

There are different approaches to calculate the Casimir force \cite{ref3,ref5,ref6}.
Perhaps, the most usual one is to calculate the Casimir
energy \cite{ref1}. This is simply the difference between
the zero point energy in the presence and the absence of the
boundaries. The Casimir energy is apparently a function of the
position of the boundaries. Once the casimir energy is evaluated,
the force is obtained easily by differentiation. This approach
can be cast into a more technical framework based upon the use of
Green function. The Green function is related to the vacuum
expectation value of the time ordered products of the field. It
is, therefore, possible to calculate the Casimir energy in terms
of the Green function at coincident arguments \cite{ref6}.
The equivalence of the sum of the zero point energy of the modes
and the vacuum expectation value of the field is straightforward.
In an alternative approach one may obtain the variation in the
electromagnetic energy when the dielectric function is varied.
The vacuum force between two semi-infinite parallel dielectrics
can be obtained in this way \cite{ref27}.

The wide application of the effect on the one hand and the recent
growth on the experimental verification of the
phenomena \cite{ref28,ref29,ref30,ref31} on the other hand,
demands some more illuminative and simpler derivation of this
effect for different quantum fields. The vacuum force is an
indication of the momentum inherent in a quantized field when the
field is in its vacuum state. We are thus led to quantize the
field for the particular geometry in question, and thereafter,
calculate the Casimir force using the energy momentum tensor.
This is the most straightforward and clear treatment which
provides both the force and the interpretation of the effect.
Apart from a few simple configurations \cite{ref32}, considerable
difficulties arise in practice when the field is to be quantized.
Any simplification in this formalism is therefore of value
whenever more complicated geometries are involved. This is
achieved by using the fluctuation dissipation theorem and Kubo's
formula \cite{ref33,ref34}. The aim of the present paper is to
develop this approach into the massive scalar field. This may be
a step forward to make a simple unified, yet comprehensive
treatment of the Casimir force in a wide variety of domains and
variously shaped configurations.

The paper is organized as follows. In Sec. II we begin with the
quantization of the massive scalar field in 1+1 space-time
dimensions in the presence of two nonpenetrating plates. The
attractive force between the two plates is obtained by the use of
the latter field operators. The formalism is developed to 3+1
space-time dimensions in Sec. III. In Sec. IV, we calculate the
vacuum force in both 1+1 and 3+1 space-time dimensions by
using the fluctuation dissipation theorem and Kubo's formula. The
generalization of the two approaches to $D+1$ space-time
dimensions are provided in Sec. V. Finally, we summarize and
discuss the main results in Sec. VI.

\section{One Dimensional Formalism}

In this section we will make use of the 1+1 space-time
dimensions. This is obviously a mathematical idealization that do
not exist in the physical world, but it is instructive. The
treatment is rather straightforward and, therefore, the basic
idea of the formalism can be explored properly.

\subsection{Field Quantization}

We begin by summarizing the elements of field quantization in an
appropriate form which are useful in the following sections. To
simplify the calculations we restrict our discussion to neutral
spin zero field ${\hat \varphi} \left (x, t \right)$ with mass
$m$. The Klein-Gordon equation governing the field of a massive
spin zero particle in one dimension is \cite{ref35}

\begin{equation}
\label{eq2} \left (\frac{\partial^2}{\partial
x^2}-\frac{1}{c^2}\frac{\partial^2}{\partial t^2}-\frac{m^2
c^2}{{\hbar}^2}\right){\hat \varphi} \left (x, t \right)=0
\end{equation}
The usual technique of Lagrangian mechanics shows that the
conjugate momentum of the field ${\hat \varphi} \left (x, t
\right)$ is

\begin{equation}
\label{eq3}{\hat \pi} \left (x, t \right)=\frac{1}{
c^2}\frac{\partial}{\partial t}{\hat \varphi} \left (x, t
\right)
\end{equation}
The quantization procedure is carried out easily by decomposing
the field ${\hat \varphi} \left (x, t \right)$ into positive and
negative frequency parts, and expanding each one in terms of an
appropriate set of mode functions $u \left(p,x \right)$ as

\begin{eqnarray}
{\hat \varphi}\left( x,t \right) &=&{\hat
\varphi}^{+}\left(x,t\right) +{\hat \varphi}
^{-}\left(x, t\right)\nonumber \\
\label{eq4} &=&\left(\frac{\hbar
c^2}{2}\right)^{1/2}\int_{-\infty}^{+\infty }\frac{dp}{\sqrt
{\omega}_p }{\hat a}\left(p,t\right)u \left(p,x \right)
+\mathrm{H.C.}.
\end{eqnarray}
where ${\hat \varphi}^{+}\left(x,t\right)$ contains all
amplitudes which vary as $\exp (-i {\omega}_p t)$ and ${\hat
\varphi}^{-}\left(x,t\right)$ contains all amplitudes which vary
as $\exp (i {\omega}_p t)$ and ${\hat
\varphi}^{-}\left(x,t\right)=\left[{\hat
\varphi}^{+}\left(x,t\right)\right]^{\dagger}$. In this
expression $p$ is the particle momentum and

\begin{equation}
\label{eq5}{\omega}_p=\left(p^2 c^2 + m^2 c^4\right)^{1/2}/\hbar
\end{equation}

We note that the use of mode function expansion has the effect
that the time dependence of annihilation operator ${\hat
a}\left(p,t\right)$ being characterized by a simple phase factor

\begin{equation}
\label{eq6}{\hat a}\left(p,t\right)={\hat a}\left(p\right)\exp
(-i {\omega}_p t)
\end{equation}
we expect that the annihilation and creation operators ${\hat
a}\left(p\right)$ and ${\hat a}^{\dagger}\left(p\right)$
fulfill the usual algebra of bosonic operators, that is

\begin{equation}
\label{eq7}\left[{\hat a}\left(p\right),{\hat
a}^{\dagger}\left(p^{\prime}\right)\right]=\delta(p-p^{\prime})
\end{equation}

The mode function $u \left(p,x \right)$ satisfies the
following differential equation

\begin{equation}
\label{eq8}\left(\frac{d^2}{dx^2}+\frac
{p^2}{{\hbar}^2}\right) u\left(p,x \right)=0
\end{equation}
This is obtained by substitution of Eq.~(\ref{eq4}) into Eq.~(\ref{eq2}), and
using Eq.~(\ref{eq6}). Regardless of the explicit form of $u \left(p,x
\right)$, it is seen from the sturm-Liouville theory that the mode
functions obtained from Eq.~(\ref{eq8}) are restricted to the
orthogonality and completeness relations if a proper boundary
condition is used and the mode functions are appropriately
normalized. The particular choice of coefficient in Eq.~(\ref{eq4}) is
made for a reason that will become evident shortly.

In order to demonstrate the full consistency of the field
quantization, one must necessary prove the equal-time canonical
commutation relation between the field operator ${\hat
\varphi}\left( x,t \right)$ and its complex conjugate ${\hat
\pi} \left (x, t \right)$

\begin{equation}
\label{eq9} \left[{\hat \varphi}\left( x,t \right),{\hat \pi}
\left (x^{\prime}, t \right)\right]=i\hbar
\delta(x-x^{\prime})
\end{equation}
This can be verified most easily by combining Eq.~(\ref{eq3}) with Eq.~(\ref{eq4}) to obtain the explicit form of ${\hat \pi} \left (x, t
\right)$, and making use of Eqs.~(7). It is seen that the
explicit form of $u \left(p,x \right)$ is not needed and the
orthogonality and completness relations of the mode functions are
the only requirements to prove Eq.~(\ref{eq9}).

\begin{figure}[!t!]
\includegraphics[width=.7\linewidth]{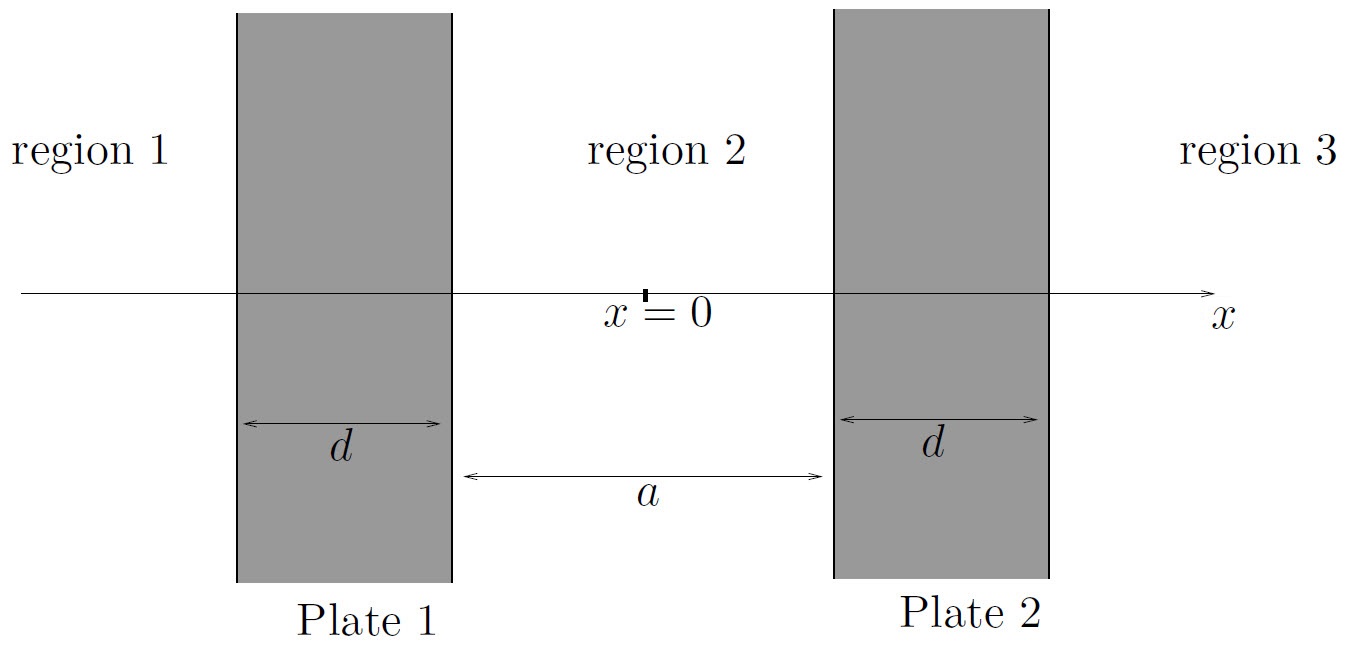}
\caption{\label{figure}
 Spatial configuration of the two plates with the geometric
parameters.}
\end{figure}

We are now in a stage to apply the presented quantization scheme
to geometrical configurations of our interest. Consider two
parallel nonpenetrating plates of thickness $d$ located in an empty space.
The $x$ axis is chosen to be perpendicular to the interfaces
with the origin at a distance $a/2$ from each plate as sketched in
Fig.~\ref{figure}. We call the left(right) hand plate as 
plate 1(2) and refer the different empty regions separated 
by the plates as region $1$, $2$ and $3$, as shown in 
Fig.~\ref{figure}.
The aim is to
obtain the field operator in the regions $2$ and $3$. It is
understood that the complete expression of the field operator in
the region $3$ is in the form of Eq.~(\ref{eq4}). In this region the
appropriate mode functions are the solution of Eq.~(\ref{eq8}) with the
boundary condition $u \left(p,x=d+a/2 \right)=0$ It is
straightforward, without elaborate algebra to show that

\begin{equation}
\label{eq12}u \left(p,x \right)=\frac{1}{\sqrt {2\pi
\hbar}}\left\{\exp\left(i p
x/\hbar\right)-\exp\left[-ip(x-a-2d)/\hbar\right]\right\}
\end{equation}
The mode functions in region $2$ are obtained from Eq.~(\ref{eq8}) with the
boundary conditions $u \left(p,x=a/2 \right)=u
\left(p,x=-a/2 \right)=0$. One can easily show that the mode
functions are also the eigenfunctions of parity operator, and
thus divided into two classes with parity $\pm$. Therefore,
\begin{equation}
\label{eq13}u_n^{-} \left(x \right)=\sqrt \frac{2}{a}{\sin}
\frac {p_n^{-}}{\hbar}x,\hspace {15mm}p_n^{-}=\frac{2n\pi
\hbar}{a}
\end{equation}
and
\begin{equation}
\label{eq14}u_n^{+} \left(x \right)=\sqrt \frac{2}{a}{\cos}
\frac {p_n^{+}}{\hbar}x,\hspace {15mm}p_n^{+}=\frac{(2n-1)\pi
\hbar}{a}
\end{equation}
where the integer $n=1,2,...$. This is the typical mode functions
of a cavity. Since $p=p_n$ takes discrete values, the
integration with respect to $p$ in Eq.~(\ref{eq4}) must be replaced with
the summation over $p_n$ or, equivalently, $n$

\begin{equation}
\label{eq15}{\hat \varphi}\left(x,t\right)=\left(\frac{\hbar
c^2}{2}\right)^{1/2}\sum_{\Omega, n}\frac{1}{\sqrt
{{\omega}_n^{\Omega} }}{\hat
a}_n^{\Omega}\left(t\right)u_n^{\Omega} \left(x \right)+H.C.
\end{equation}
where $\Omega=(\pm)$ and ${\omega}_n^{\Omega}$ is $\omega_p$ for
$p=p_n^{\Omega}$. We note that in this region the Dirac delta
function in the right hand side of Eq.~(\ref{eq7}) must be replaced by
Kronecker delta functions
\begin{equation}
\label{eq16}\left[{\hat a}_n^{\Omega},{\hat
a}_{n^{\prime}}^{{\Omega}^{\prime} \dagger}\right]={\delta}_{n
n^{\prime}}{\delta}_{\Omega \Omega^{\prime}}.
\end{equation}

\subsection{Casimir Force}
The statement of conservation of linear momentum for a classical
field shows that the flow per unit area of linear momentum across
the surface $S$ into volume $V$ is given by ${\bf T} \cdot {\bf
\tilde n}$, where ${\bf \tilde n}$ is the unit outward normal
vector at the surface $S$. The three dimensional tensor ${\bf T}$
is the stress tensor associated with the classical field. It is
in fact the space-space components of the four dimensional
symmetric mixed canonical energy momentum tensor. The explicit
form of ${\bf T}$ for the massive scalar field is

\begin{equation}
\label{eq18}T_{\alpha \beta}=-\frac {\partial\varphi}{\partial
x_{\alpha}}\frac {\partial\varphi}{\partial
x_{\beta}}+\frac{1}{2}\left[(\mbox{\boldmath
$\nabla$}\varphi)\cdot (\mbox{\boldmath $\nabla$}\varphi)-\frac
{1}{c^2}\left(\frac{\partial\varphi}{\partial t}\right)^{2}+\frac
{m^2 c^2}{\hbar^2}\varphi^2 \right]\delta_{\alpha \beta}
\end{equation}
where $\alpha, \beta=1,2,3$. 

To describe the force in the
quantum domain, we must replace the classical field by the
corresponding field operator. The expectation value of the force
operator evaluated for a given state of the field (the vacuum state in our case) will yield the
force.

Let us now calculate the Casimir force acting on plate $2$ in
Fig.~\ref{figure} Recalling the symmetry of the present configuration, it is
evident that the force per unit area acting on the
exterior(interior) interface of plate $2$ is given by
${T_{11}}$(-${T_{11}}$). It is seen from Eq.~(\ref{eq18}) that the
explicit form of ${T_{11}}$ in 1+1 space-time dimensions is

\begin{equation}
\label{eq20}T_{11}=-\frac{1}{2}\left[\left(\frac
{\partial\varphi}{\partial x}\right)^2+\frac
{1}{c^2}\left(\frac{\partial\varphi}{\partial t}\right)^{2}-\frac
{m^2 c^2}{\hbar^2}\varphi^2 \right]
\end{equation}
To simplify the force expression it is convenient to separate the
field operator into positive and negative frequency parts. We,
therefore, find that

\begin{equation}
\label{eq21}F=\mp\frac{1}{2}\langle 0|\left(\frac
{\partial\varphi^+}{\partial x}\right)\left(\frac
{\partial\varphi^-}{\partial x}\right)+\frac
{1}{c^2}\left(\frac{\partial\varphi^+}{\partial
t}\right)\left(\frac{\partial\varphi^-}{\partial t}\right)-\frac
{m^2 c^2}{\hbar^2}\varphi^+ \varphi^-|0\rangle
\end{equation}
where the upper(lower) sign holds for the force per unit area
acting on the exterior(interior) interface of plate $2$ and $|0
\rangle$ refers to the vacuum state of the field. Note that the
terms in which the annihilation operator acts directly on the
vacuum state of the field have been set to zero.

The vacuum radiation pressure experienced by the exterior
interface of plate 2 as well as its interior interface due to the vacuum field in the regions $x>a/2 +d$  and 
 $-a/2<x<a/2$ respectively, are
obtained by the substitution of Eqs.~(\ref{eq4}) and (\ref{eq5}) into Eq.~(\ref{eq21}). These calculations are not
provided here for the sake of brevity. The net force per unit area acting on plate 2 can 
be written down as
\begin{equation}
\label{eq27}F=\frac{c^2}{2a \hbar}\sum_{
n=1}^{\infty}\frac{{p_n}^2}{{\omega_n}}-\frac{c^2}{2 \pi
\hbar^2}\int_{0}^{+\infty }dp\frac{p^2}{{\omega}_p}\,.
\end{equation}
It is convenient to introduce $n=\frac{a}{\pi \hbar}p$ as the
variable of integration and rewrite Eq.~(\ref{eq27}) in the form of
\begin{equation}
\label{eq28}F=\frac{ \pi^2 \hbar^2 c}{2a^3}\left[\sum_{
n=0}^{\infty}f(n)-\int_{0}^{+\infty }dn f(n)\right]
\end{equation}
where

\begin{equation}
\label{eq29}f(n)=\frac{n^2}{\sqrt {\left(\frac{n \pi
\hbar}{a}\right)^2+m^2 c^2}}
\end{equation}
Note that the term $n=0$ in Eq.~(\ref{eq28}) is zero and thus Eq.~(\ref{eq27}) and
Eq.~(\ref{eq28}) are identical. Using the poisson's sum formula \cite{ref3,ref351}, the
summation of $f(n)$ over $n$ can be treated as the summation of
the Fourier cosine transform of $f$ over $n$, that is

\begin{equation}
\label{eq30}F=\frac{\pi^2 \hbar^2 c}{2a^3}\left[\sum_{
n=-\infty}^{+\infty}\int_{0}^{+\infty }dx f(x) \cos (2\pi nx
)-\int_{0}^{+\infty }dn f(n)\right].
\end{equation}
It is seen that in the squared brackets in the right hand side of Eq.~(\ref{eq30}), the first term for $n=0$ and the second term cancel one another and we
find that
\begin{equation}
\label{eq31}F=-\frac{{\hbar}^2 c}{4 a^3}\sum_{
n=1}^{\infty}\frac{\partial^2}{\partial n^2}\int_{0}^{+\infty }dx
\frac{ \cos (2\pi nx )}{\sqrt {\left(\frac{x \pi
\hbar}{a}\right)^2+m^2 c^2}}\,.
\end{equation}
The integral representation of the modified Bessel function $K_0$
can be used for integration over $x$ and the recurrence relations of modified Bessel functions can be
employed for differentiation in Eq.~(\ref{eq31}) \cite{ref36}.
 We finally obtain the force expression as

\begin{equation}
\label{eq33}F=-\frac{m^2 c^3}{\pi \hbar}\sum_{
n=1}^{\infty}\left[K_2 (n \xi)-\frac{1}{n \xi}K_1(n \xi)\right].
\end{equation}
Employing the limiting forms of $K_1$ and $K_2$ for small
argument, one can easily show that the Casimir force for a neutral
massless scalar field in 1+1 dimensions is

\begin{equation}
\label{eq34}F=-\frac{\pi \hbar c}{24 a^2}
\end{equation}
This is half of the similar expression for the electromagnetic
field. This is understood in terms of the two states of
polarization of the electromagnetic field \cite{ref3}.

\section{Three Dimensional Formalism}

A great variety of problems of practical and theoretical interest
are in three space dimensions. The development of the present
formalism to 3+1 space-time dimensions is, therefore,
important. The purpose of this section is to set 3+1
dimensional formalism in a context to establish a close contact
with the results already obtained in 1+1 dimensions. This
allows us to avoid any duplication for similar calculations.

\subsection{Field Quantization}

The dynamical behaviour of a neutral spin zero field $\hat
\varphi ({\bf r}, t)$ with mass $m$ is obtained from Klein-Gordon
equation.

\begin{equation}
\label{eq35}\left({\mbox{\boldmath $\nabla$}^2}
-\frac{1}{c^2}\frac{\partial^2}{\partial t^2}-\frac {m^2
c^2}{\hbar^2}\right){\hat \varphi ({\bf r}, t)}=0
\end{equation}
where ${\bf r}=(x_1,x_2,x_3)$. We usually work with a complete
continuum of plane wave mode functions in which the ${\bf p}$
vector are not restricted to the discrete spectrum. As in the
1+1 dimensional formulation it is useful to decompose the field
operator into positive and negative frequency parts $\hat
\varphi^{\pm} ({\bf r}, t)$. Expanding each part in terms of an
appropriate set of mode functions $u({\bf p},{\bf r})$, we find
that

\begin{equation}
\label{eq36}{\hat \varphi}\left( {\bf r},t \right)=
\left(\frac{\hbar c^2}{2}\right)^{1/2}\int \frac{d^3 {\bf
p}}{\sqrt {\omega}_p }{\hat a}\left({\bf p },t\right)u \left({\bf
p},{\bf r} \right) +\mathrm{H.c.}.
\end{equation}
Comparing Eqs.~(\ref{eq35}) and (\ref{eq36}) with their corresponding 1+1
dimensional forms Eq.~(\ref{eq2}) and (\ref{eq4}), it is seen that the
annihilation operator ${\hat a}\left({\bf p },t\right)$ has the
same time dependence as in Eq.~(\ref{eq6}), and possesses the bosonic
commutation relation

\begin{equation}
\label{eq37}\left[{\hat a}\left({\bf p}\right),{\hat
a}^{\dagger}\left({\bf p }^{\prime}\right)\right]=\delta({\bf p
}-{\bf p}^{\prime})
\end{equation}
It can be verified from substitution of Eq.~(\ref{eq36}) into Eq.~(\ref{eq35})
that the mode functions are obtained from the differential
equation

\begin{equation}
\label{eq38}\left({\mbox{\boldmath $\nabla$}^2} +
\frac{p^2}{\hbar^2}\right)u({\bf p},{\bf r})=0
\end{equation}
For some purposes which become evident shortly, it is useful to
cast the mode functions for free space in the absence of any
boundary surface into some what different form

\begin{equation}
\label{eq39}u\left({\bf p, {\bf r}}\right)=\frac{1}{2 \pi
\hbar}\exp(i{\bf p}_{\parallel}\cdot {\bf
x}_{\parallel}/\hbar)u(p_1, x_1)
\end{equation}
where ${\bf x}_{\parallel}=(x_2,x_3)$, ${\bf p}_{\parallel}=(p_2,
p_3)$ and $u(p_1, x_1)$ is obtained by solving Eq.~(\ref{eq8}) and replacing $x$ and $p$ with $x_1$ and $p_1$, respectively. 

We now consider briefly the field quantization in the region $3$
of Fig.~\ref{figure} in three space dimensions. We note that the $x_2x_3$
plane is assumed to be parallel to the plane interfaces of the
two plates. It is convenient to retain the forms of Eqs.~(\ref{eq36}) and
(\ref{eq39}) for the field operator and the mode functions. This means
that the explicit form of $u(p_1, x_1)$ is now given by Eq.~(\ref{eq12}).

A more complicated form for the field operator appears in the
region $2$. In this case we have to work with a complete plane
wave mode functions in which the $p_1$ component of the ${\bf p}$
vectors are restricted to the discrete spectrum, while their
${\bf p_{\parallel}}$ are not. This indicates that the proper
form for the field operator is

\begin{equation}
\label{eq40}{\hat \varphi}\left({\bf r
},t\right)=\left(\frac{\hbar c^2}{2}\right)^{1/2}\int d^2 {\bf
p}_{\parallel} \sum_{\Omega, n}\frac{1}{\sqrt {{\omega}_n^{\Omega}
}}{\hat a}_n^{\Omega}\left({\bf p}_{\parallel},
t\right)u_n^{\Omega} \left({\bf p}_{\parallel}, {\bf r}
\right)+H.c.
\end{equation}
where

\begin{equation}
\label{eq41}u_n^{\Omega}\left({{\bf p}_{\parallel}, {\bf
r}}\right)=\frac{1}{2 \pi \hbar}\exp(i{\bf p}_{\parallel}\cdot
{\bf x}_{\parallel}/\hbar)u_n ^{\Omega}(x_1)
\end{equation}
Note that $u_n ^{\Omega}(x_1)$ are given by Eqs.~(\ref{eq13}) 
and (\ref{eq14}). 
Recalling the typical algebra of bosonic operators inside the
cavity in one dimension, we are led to consider the following
algebra in the present case

\begin{equation}
\label{eq42}\left[{\hat a}_n^{\Omega}({\bf p}_{\parallel}),{\hat
a}_{n^{\prime}}^{{\Omega}^{\prime} \dagger}({\bf p
}_{\parallel}^{\prime})\right]={\delta}_{n
n^{\prime}}{\delta}_{\Omega \Omega^{\prime}}\delta({\bf
p}_{\parallel}-{\bf p}_{\parallel}^{\prime})
\end{equation}
The question such as the orthogonality and completeness relations
of the mode functions in each region in three dimensions are
easily verified in terms of Eq.~(\ref{eq39}) and the corresponding one
dimensional problem.

\subsection{Casimir Force}

Having established the appropriate forms for the field operator
in regions $2$ and $3$ in 3+1 dimensions, we can now seek to
establish the force expression. Decomposing the field into
positive and negative frequency parts, it is straightforward to
show that the force per unit area acting on the exterior(interior)
interface of plate $2$ is

\begin{equation}
\label{eq43}F=\mp\frac{1}{2}\langle 0|\left(\frac
{\partial\varphi^+}{\partial x_1}\right)\left(\frac
{\partial\varphi^-}{\partial x_1}\right)-\left(\frac
{\partial\varphi^+}{\partial x_{\parallel}}\right)\left(\frac
{\partial\varphi^-}{\partial x_{\parallel}}\right)+\frac
{1}{c^2}\left(\frac{\partial\varphi^+}{\partial
t}\right)\left(\frac{\partial\varphi^-}{\partial t}\right)-\frac
{m^2 c^2}{\hbar^2}\varphi^+ \varphi^-|0\rangle
\end{equation}
with the sign convention of Eq.~(\ref{eq21}).

We are now in a position to evaluate the force on the exterior
and interior interfaces of plate $2$ and express the net force
as
\begin{equation}
\label{eq49}F=\frac{1}{(2 \pi \hbar)^2} \int d^2 {\bf
p}_{\parallel} \left\{ \frac{ \pi^2 \hbar^2 c}{2a^3}\left[\sum_{
n=0}^{\infty}f(n, p_{\parallel})-\int_{0}^{+\infty }dn f(n,
p_{\parallel})\right]\right\}
\end{equation}
where
\begin{equation}
\label{eq50}f(n, p_{\parallel})=\frac{n^2}{\sqrt {\left(\frac{n
\pi \hbar}{a}\right)^2+p_{\parallel}^2+m^2 c^2}}
\end{equation}
Though the two terms in Eq.~(\ref{eq49}) diverge, it
is straightforward with the help of Poisson's sum formula to
obtain a finite net force acting on plate $2$. Following the
steps from Eq.~(\ref{eq28}) to Eq.~(\ref{eq31}), the force can be written as

\begin{equation}
\label{eq51}F=\frac{1}{(2 \pi \hbar)^2} \int d^2 {\bf
p}_{\parallel} \left\{ -\frac{{\hbar}^2 c}{4 a^3}\sum_{
n=1}^{\infty}\frac{\partial^2}{\partial n^2}\int_{0}^{+\infty }dx
\frac{ \cos (2\pi nx )}{\sqrt {\left(\frac{x \pi
\hbar}{a}\right)^2+p_{\parallel}^2+m^2 c^2}}\right\}
\end{equation}
Employing the integral representation of the modified Bessel function $K_0$
 and using the recurrence relations of modified Bessel functions for differentiation in Eq.~(\ref{eq51}) \cite{ref36}, the force can be obtained as 

\begin{equation}
\label{eq54}F=-\frac{m^4 c^5}{2 \pi^2 \hbar^3}\sum_{
n=1}^{\infty} \frac{1}{n\xi}\left[K_3 (n \xi)-\frac{1}{n \xi}K_2(n
\xi)\right]
\end{equation}
The Casimir force for a neutral massless scalar field in 3+1
space-time dimensions is obtained by using the limiting forms of
$K_2$ and $K_3$ for small argument

\begin{equation}
\label{eq55}F=-\frac{\pi^2 \hbar c}{480 a^4}
\end{equation}
As in the one dimensional formalism, Eq.~(\ref{eq55}) is half of the
similar expression for massless vector field.

We thus introduce a formal approach for the calculation of the
Casimir force for massless scalar field. It is, however, useful
to cast this formalism into a some what different form.

\section{Casimir Effect and Fluctuation Dissipation Theorem}

It is seen that in this formalism one should necessarily go
through the process of the field quantization for the particular
geometry in question and then calculate the Casimir force using
the stress tensor. However, the field quantization encounters considerable difficulties,
except for a few simple geometries. Any
simplification in this formulation to avoid the use of explicit
form of the field operator is therefore appreciated whenever more
complicated geometries are involved.

It is convenient to separate the field operator into positive and
negative frequency parts in the usual way

\begin{eqnarray}
{\hat \varphi}\left({\bf r}, t\right) &=&{\hat
\varphi}^{+}\left({\bf r},
t\right)+{\hat \varphi}^{-}\left({\bf r}, t\right)\nonumber \\
\label{eq56} &=&\frac 1{\sqrt{2\pi }}\int_{mc^2/\hbar}^{+\infty
}d\omega_p { \hat \varphi} ^{+}\left( {\bf r},\omega_p \right)
\exp \left( -i\omega_p t\right) +\mathrm{H.c.},
\end{eqnarray}
It is clear that the positive and negative frequency parts in the
integrand involve only the annihilation and creation operators,
respectively. Note that the lower limit of integration in 
Eq.~(\ref{eq56}) is the condition which implies by Eq.~(\ref{eq5}).

The quantized field correlation function can be related to the
imaginary part of the field Green function in the frequency
domain $G({\bf r}, {\bf r}^{\prime}, \omega_p)$ by using
fluctuation dissipation theorem and Kubo's formula \cite{ref37}

\begin{equation}
\label{eq57} \langle 0|{ \hat \varphi }^{+}\left({\bf r,
\omega_p}\right){ \hat \varphi }^{-}\left({\bf r^{\prime},
\omega_p^{\prime}}\right)|0\rangle= 4 \hbar
{\mathrm{Im}}G\left({\bf r},{\bf r}^{\prime},\omega_p\right)\delta
\left( \omega_p -\omega_p^{\prime}\right).
\end{equation}
Note that the right hand side of Eq.~(\ref{eq57}) is half of the similar
expression for the electromagnetic field. This arises from the
two states of polarization of the electromagnetic field. The
whole idea in this approach is based on the understanding that we
can express Eq.~(\ref{eq21}) or Eq.~(\ref{eq43}) in terms of the field correlation
function such that to use Eq.~(\ref{eq57}) for the calculation of the
Casimir force. It is useful to treat the problem for both 1+1
and 3+1 space-dimensions separately.

\subsection{Casimir Force in 1+1 dimensions}

It is seen that in order to calculate the force per unit area
acting on plate $2$, we need the explicit form of the coordinate
space Green function for the case where both the source and
observation points are on either side of the plate. One can
easily show from Eq.~(\ref{eq2}) that the Fourier-time transformed Green
function in 1+1 dimensions is determined by the solution of

\begin{equation}
\label{eq58}\left(\frac{d^2}{dx^2}+\frac
{p^2}{{\hbar}^2}\right) G\left(x, x^{\prime}, \omega_p
\right)=-\delta (x-x^{\prime})
\end{equation}
The boundary conditions on $G\left(x, x^{\prime}, \omega_p
\right)$ are divided into two types. The first are the boundary
conditions at $\pm\infty$, which are easily imposed by assuming
outgoing travelling waves. The second are those at the plane
interfaces of the plates, which are governed by the boundary
conditions on the field. The vanishing of the field operator on
the plates entails that the Green function should necessarily
vanish when the observation point $x$ is on the interface of the
plates. The details of these calculation are omitted here for the
sake of brevity.

The coordinate space Green function is

\begin{equation}
\label{eq59}G(x, {x}^{\prime}, \omega_p)=\frac{i \hbar}{2
p}\left\{\exp[ip|x-x^{\prime}|/
\hbar]-\exp[ip(x+x^{\prime}-a-2d)/\hbar]\right\}
\end{equation}
where the two points $x$ and $x^{\prime}$ are both within
region $3$. The two terms in Eq.~(\ref{eq59}) are typical of the
semi-infinite geometry. The bulk part term with argument
$|x-x^{\prime}|$ is associated with the direct communication
between $x$ and $x^{\prime}$, while the other term corresponds
to the communication between $x$ and $x^{\prime}$ via
reflection in the exterior interface of of plate $2$.

If $x$ and $x^{\prime}$ are both within region $3$, the
appropriate Green function in the coordinate space is
\begin{eqnarray}
\label{eq60}
&&G(x, {x}^{\prime}, \omega_p)=
\frac{i\hbar}{2p}\exp(ip|x-x^{\prime}|/\hbar)-\frac{i\hbar}{2p} \frac{\exp(2ipa/\hbar)}{1-\exp(2ipa/\hbar)}
\big\{\exp[-ip(x+x^{\prime}+a)/\hbar]
\nonumber \\
&&\hspace{1.5in}+\exp[ip(x+x^{\prime}-a)/\hbar]-\exp[-ip(x-x^{\prime})/\hbar]-\exp[ip(x-x^{\prime})/\hbar]\big\}.
\end{eqnarray}
The structure of this Green function is typical of the cavity
systems. The first term, the bulk part, corresponds to the direct
communication between $x$ and $x^{\prime}$, while the rest is associated with the communication between $x$ and
$x^{\prime}$ via reflections from the cavity walls. Note that in
the derivation of Eqs.~(\ref{eq59}) and (\ref{eq60}), the boundary conditions
requires that these two Green function vanish on the exterior and
interior interface of plate $2$, respectively.

Having found the coordinate space Green function on either side
of plate $2$, the evaluation of the Casimir force acting on plate
$2$ is straightforward by means of Eqs.~(\ref{eq21}) and
(\ref{eq57}). It seems
clear, in view of the typical behaviour of the Green function on
the interfaces of plate $2$, that the only nonvanishing term of
the force expression in 1+1 dimensions is

\begin{equation}
\label{eq61}F=\mp\frac{1}{2}\langle 0|\frac {\partial}{\partial
x} \hat \varphi^+(x,t) \frac {\partial}{\partial x}
 \hat \varphi^-(x,t)|0\rangle
\end{equation}
This means that only the spatial derivatives of the field
correlation function is not zero on the interfaces. Substitution
of the field operator for 1+1 dimensions from Eq.~(\ref{eq56}) into Eq.~(\ref{eq61}) yields

\begin{equation}
\label{eq62}F=\mp\frac{1}{4\pi}\int_{mc^2/\hbar}^{+\infty}d\omega_p
\int_{mc^2/\hbar}^{+\infty}d\omega_p^{\prime}\left[
\frac{\partial}{\partial x}\frac{\partial}{\partial
x^{\prime}}\langle 0|  \hat \varphi^{+}(x,\omega_p) \hat
\varphi^{-}(x^{\prime}, \omega_p^{\prime}) |0\rangle \right
]_{x=x^{\prime}}\exp[-i(\omega_p -\omega_p^{\prime})t]
\end{equation}
Employing Eq.~(\ref{eq57}), the force expression can be rewritten as

\begin{equation}
\label{eq63}
F=\mp\frac{\hbar}{2\pi}\int_{mc^2/\hbar}^{\infty}d\omega_p
\left[ \frac{\partial}{\partial x}\frac{\partial}{\partial
x^{\prime}} {\mathrm{Im}}G\left(x,
x^{\prime},\omega_p\right) \right ]_{x=x^{\prime}}
\end{equation}
The Casimir force acting on plate $2$ is evaluated by taking into
account the vacuum radiation pressure on both side of this plate.
Using Eq.~(\ref{eq63}), the net force per unit area is

\begin{equation}
\label{eq64}F=\frac{\hbar}{2\pi}\int_{mc^2/\hbar}^{\infty}d\omega_p
\left\{\left[ \frac{\partial}{\partial
x}\frac{\partial}{\partial x^{\prime}}
{\mathrm{Im}}G\left(x, x^{\prime},\omega_p\right) \right
]_{x=x^{\prime}=a/2}-\left[ \frac{\partial}{\partial
x}\frac{\partial}{\partial x^{\prime}}
{\mathrm{Im}}G\left(x, x^{\prime},\omega_p\right) \right
]_{x=x^{\prime}=a/2+d}\right\}
\end{equation}
The coordinate space Green function needed for substitution into
the first and second terms of Eq.~(\ref{eq64}) are given by Eqs.~(\ref{eq60}) and
(\ref{eq59}), respectively. The bulk part contributions, which are
identical in both Eqs.~(\ref{eq59}) and (\ref{eq60}), cancel each other in Eq.~(\ref{eq64}).
 Furthermore, on expanding the prefactor of the square
brackets in Eq.~(\ref{eq60}), one can easily show, by an appropriate
manipulation of the summation indices, that the four terms of 
Eqs.~(\ref{eq60})
 can be written as summation over $n$, where in the first term
$n$ varies from $0$ to $\infty$, while in the other three terms 
it
varies from $1$ to $\infty$. The reflection term in Eq.~(\ref{eq59}) on
the exterior interface of plate $2$ is identical with the first
term in Eq.~(\ref{eq60}) with $n=0$ on the interior interface. These two
terms cancel one another as well. The result after some algebra
can be rewritten as

\begin{equation}
\label{eq65}F=\frac{1}{\pi}\int_{mc^2/\hbar}^{\infty}d\omega_p p
\sum_{n=1}^{\infty} \cos(2np a/ \hbar)
\end{equation}
It is more convenient to choose $p$ as the variable of
integration. Therefore

\begin{equation}
\label{eq66}F=\frac{c}{\pi
\hbar}\sum_{n=1}^{\infty}\int_{0}^{\infty}dp \frac{{p}^2
\cos(2np a/ \hbar)}{\sqrt{{p}^2+m^2 c^2}}
\end{equation}
The integration in Eq.~(\ref{eq66}) is all that need be done. To
establish contact with the result already obtained, given by 
Eq.~(\ref{eq31}), it is adequate to introduce $x=ap/ \pi \hbar$ as the
variable of integration. The treatment is straightforward and the
force expression can, therefore, be simplified as Eq.~(\ref{eq33}).

\subsection{Casimir Force in 3+1 dimensions}
As in 1+1 dimensions, we need the explicit form of the
coordinate space Green function for the case where both the
source and observation points ${\bf r}$ and ${\bf r}^{\prime}$ are
within the gap between the two plates as well as where both ${\bf
r}$ and ${\bf r}^{\prime}$ are in region $3$. One can easily
begin with the usual definition of the Fourier time transformed
Green function along with the use of Eq.~(\ref{eq35}) to show that the response function satisfies the following differential equation
\begin{equation}
\label{eq67}\left({\mbox{\boldmath $\nabla$}^2}
+\frac{p^2}{\hbar^2}\right)G({\bf r}, {\bf r}^{\prime},
\omega_p)=-\delta({\bf r}-{\bf r}^{\prime}).
\end{equation}
The symmetry of the present configuration enables us to convert
this partial differential equation into an ordinary differential
equation in variable $x_1$. This is obtained by expressing the
response function in terms of its Fourier transform as

\begin{equation}
\label{eq68} G({\bf r}, {\bf r}^{\prime}, \omega_p)=\frac{1}{(2
\pi \hbar)^2} \int d^2 {\bf p}_{\parallel} G({\bf p}_{\parallel},
x_1, x_1^{\prime}, \omega_p) \exp[i{\bf p}_{\parallel}\cdot ({\bf
x}_{\parallel}-{\bf x}_{\parallel}^{\prime})/ \hbar]
\end{equation}
Substitution of Eq.~(\ref{eq68}) into Eq.~(\ref{eq67}) shows that $G({\bf
p}_{\parallel}, x_1, x_1^{\prime}, \omega_p)$ satisfies

\begin{equation}
\label{eq69}\left(\frac{\partial^2}{\partial x_1^2}
+\frac{{p_1}^2}{\hbar^2}\right)G({\bf p}_{\parallel}, x_1,
x_1^{\prime}, \omega_p)=-\delta(x_1-x_1^{\prime})
\end{equation}
It is obvious that the boundary conditions on Eq.~(\ref{eq69}) are the
same as those applies on Eq.~(\ref{eq58}). It, therefore, follows that
the solutions of Eq.~(\ref{eq69}) on the right and left side of plate $2$
are in the form of Eqs.~(\ref{eq59}) and (\ref{eq60}), respectively.

The coordinate space Green function corresponding to Eq.~(\ref{eq59}) can
be written in the form of
\begin{equation}
\label{eq70} G({\bf r}, {\bf r}^{\prime}, \omega_p)={\cal G}({\bf
r}_{0}^{rel}, \omega_p)-{\cal G}({\bf r}_{1}^{rel}, \omega_p)
\end{equation}
where the first term corresponds to the bulk part and defined as
\begin{equation}
\label{eq71}{\cal G}({\bf r}_{0}^{rel}, \omega_p)=\frac{i}{8
\pi^2 \hbar}\int \frac{d^2 {\bf
p}_{\parallel}}{p_1}\exp[ip_1|x_1-x_1^{\prime}|/ \hbar] \exp[i{\bf
p}_{\parallel}\cdot ({\bf x}_{\parallel}-{\bf
x}_{\parallel}^{\prime})/ \hbar]
\end{equation}
where

\begin{equation}
\label{eq72}{\bf r}_{0}^{rel}=(x_1-x_1^{\prime}){\bf \hat {\bf x_1
}}+{\bf x}_{\parallel}^{rel}, \hspace {15mm} {\bf
x}_{\parallel}^{rel}={\bf x}_{\parallel}-{\bf
x}_{\parallel}^{\prime}
\end{equation}
The second term in Eq.~(\ref{eq70}) is associated with the surface term
having the following form
\begin{equation}
\label{eq73}{\cal G}({\bf r}_{1}^{rel}, \omega_p)=\frac{i}{8 \pi^2
\hbar}\int \frac{d^2 {\bf
p}_{\parallel}}{p_1}\exp[ip_1(x_1+x_1^{\prime}-a-2d)/ \hbar]
\exp[i{\bf p}_{\parallel}\cdot ({\bf x}_{\parallel}-{\bf
x}_{\parallel}^{\prime})/ \hbar]
\end{equation}
where
\begin{equation}
\label{eq74}{\bf r}_{1}^{rel}=(x_1+x_1^{\prime}-a-2d){\bf \hat
x_1}+{\bf x}_{\parallel}^{rel}.
\end{equation}

The integration in Eqs.~(\ref{eq71}) and (\ref{eq73}) 
can be done in a
straightforward manner to find the coordinate space Green
function. The treatment is lengthy and rather boring and
conventional. We actually look for a plain, and not elaborate,
argument to obtain the coordinate space response function. The
bulk part can be thought of as the 
Green function in the absence of
any boundary surfaces. In this sense 
${\cal G}({\bf r}_{0}^{rel},
\omega_p)$ is the Green function of Helmholtz differential
equation, given by Eq.~(\ref{eq67}), in the absence of any boundary
surfaces. This is a well known response function which we write
it down in the following form for the reason that will be clear
shortly
\begin{equation}
\label{eq75} {\cal G}({\bf r}_{i}^{rel}, \omega_p)=\frac{1}{4 \pi
{\bf r}_{i}^{rel}}\exp(ip{\bf r}_{i}^{rel}/ \hbar), \hspace
{15mm}i=0,1,...
\end{equation}
where ${\bf r}_{i}^{rel}$ for $i=0$ denotes the position of the
observation point relative to the source point. From the
similarity of Eq.~(\ref{eq73}) with Eq.~(\ref{eq71}), it is seen that the response
function ${\cal G}({\bf r}_{1}^{rel}, \omega_p)$ can also be
written in the form of Eq.~(\ref{eq75}) where the position of the
observation points relative to the source point is now defined as
${\bf r}_{1}^{rel}$, given by Eq.~(\ref{eq74}). The structure of Eq.~(\ref{eq70})
is typical of a semi-infinite free space Green function. The
surface part, the second term, that corresponds to the
communication between the points ${\bf r}$ and ${\bf r}^{\prime}$
via reflection in the plane interface is associated with the
image source.

We are now in a position to adopt this mathematical treatment for
the calculation of the coordinate space Green function
corresponding to Eq.~(\ref{eq60}). In fact the expansion of the prefactor
of the square brackets in Eq.~(\ref{eq60}) prepares the ground for the
use of latter technique in a straightforward manner. This allows
one to cast the response function into the standard form as

\begin{equation}
\label{eq76} G({\bf r}, {\bf r}^{\prime}, \omega_p)={\cal G}({\bf
r}_{0}^{rel}, \omega_p)-\sum _{n=1}^{\infty}\left[{\cal G}({\bf
r}_{2}^{rel}, \omega_p)+{\cal G}({\bf r}_{3}^{rel},
\omega_p)-{\cal G}({\bf r}_{4}^{rel}, \omega_p)-{\cal G}({\bf
r}_{5}^{rel}, \omega_p)\right]
\end{equation}
where the position of the observation point relative to the
source point ${\bf r}_{i}^{rel}$ for $i=2,3,4,5$ have the same
${\bf x}_{\parallel}^{rel}$ while their ${x}_{i}^{rel}$
coordinates are

\begin{eqnarray}
&&{x}_{2}^{rel}=(2n-1)a-(x_1+x_1^{\prime})\hspace {20mm}
{x}_{4}^{rel}=2na-(x_1-x_1^{\prime})\nonumber \\
\label{eq77}&&{x}_{3}^{rel}=(2n-1)a+(x_1+x_1^{\prime})\hspace
{20mm} {x}_{5}^{rel}=2na+(x_1-x_1^{\prime})
\end{eqnarray}
The structure of Eq.~(\ref{eq76}) is typical of a cavity geometry made up
of two perfectly nonpenetrating walls. The first term is the bulk
part, while the other terms correspond to the communication
between the two points ${\bf r}$ and ${\bf r}^{\prime}$ via a
series of infinite reflections in the cavity walls.

The coordinate space Green functions Eqs.~(\ref{eq70}) and (\ref{eq76}) enable
us, at least in principle, to calculate the Casimir force acting
on plate $2$ by the help of fluctuation dissipation theorem. The
obvious and fundamental feature of the correlation function,
provided by Eq.~(\ref{eq57}), is that only its derivatives with respect
to $x_1$ or $x_1^{\prime}$ may possibly give a nonvanishing value
on either side of plate 2. This means that the force expression in
3+1 dimensions can be rewritten in the form of

\begin{equation}
\label{eq78}F=\mp\frac{\hbar}{2\pi}\int_{mc^2/\hbar}^{\infty}d\omega_p
\left[ \frac{\partial}{\partial x_1}\frac{\partial}{\partial
x_1^{\prime}} {\mathrm{Im}}G\left({\bf r}, {\bf r
}^{\prime},\omega_p\right) \right ]_{{\bf r}={\bf r}^{\prime}}
\end{equation}
on the either side of plate $2$. This is obtained by combining
Eqs.~(\ref{eq43}), (\ref{eq56}) and (\ref{eq57}). In this way, we can calculate the net
force per unit area from the analogue of Eq.~(\ref{eq64}) for 3+1
dimensions

\begin{equation}
\label{eq79}F=\frac{\hbar}{2\pi}\int_{mc^2/\hbar}^{\infty}d\omega_p
\left\{\left[ \frac{\partial}{\partial
x_1}\frac{\partial}{\partial x_1^{\prime}}
{\mathrm{Im}}G\left({\bf r}, {\bf r }^{\prime},\omega_p\right)
\right ]_{\begin{array}{c}{\bf r}={\bf
r}^{\prime}\\x_1=x_1^{\prime}=a/2\end{array}}-\left[
\frac{\partial}{\partial x_1}\frac{\partial}{\partial
x_1^{\prime}} {\mathrm{Im}}G\left({\bf r}, {\bf
r}^{\prime},\omega_p\right) \right ]_{\begin{array}{c}{\bf r}={\bf
r}^{\prime}\\x_1=x_1^{\prime}=a/2+d \end{array}}\right\}
\end{equation}
It should be obvious that Eq.~(\ref{eq76}) and (\ref{eq70}) must be substituted
into the first and second term of Eq.~(\ref{eq79}), respectively. It is
seen that the bulk part contributions cancel each other.
Furthermore, the first term in Eq.~(\ref{eq76}) whose relative position
vector is ${\bf r}_{2}^{rel}$ with $n=1$ on the interior
interface of plate $2$ is identical with the reflection term in
Eq.~(\ref{eq70}) on the exterior interface of the plate. These two terms
also cancel each other. Changing the index $n$ in the first term
as $n\rightarrow n-1$, along with taking into account that
$\partial/\partial x_1=\partial/\partial x_1^{\prime}$ in the
second and third terms of Eq.~(\ref{eq76}), while $\partial/\partial
x_1=-\partial/\partial x_1^{\prime}$ in the forth and fifth terms,
it is not difficult after some rearrangement to show that the net
force takes the simple form

\begin{equation}
\label{eq80}F=-\frac{\hbar}{2\pi^2}{\mathrm{Im}}\sum_{n=1}^{\infty}
\int_{mc^2/\hbar}^{\infty}d\omega_p\left[\frac{(ip/\hbar)^2}{(2na)}
-2\frac{(ip/\hbar)}{(2na)^2}+\frac{2}{(2na)^3}\right]\exp[ip(2na)/\hbar]
\end{equation}
Taking the imaginary part of the summation along with changing
the variable of integration from $\omega_p$ to $p$, one can
easily show that

\begin{equation}
\label{eq81}F=\frac{\hbar c}{32\pi^2 a^4 }\sum_{n=1}^{\infty}
\left(\frac{1}{n}\frac{\partial^2}{\partial
n^2}-\frac{2}{n^2}\frac{\partial}{\partial
n}+\frac{2}{n^3}\right)\frac{\partial}{\partial
n}\int_{0}^{\infty}dp \frac{\cos(2npa/ \hbar)}{\sqrt{p^2+m^2 c^2}}
\end{equation}
The integration in Eq.~(\ref{eq81}) can be done easily by using the
integral representation of the modified Bessel function $K_0$.
The result is the same as Eq.~(\ref{eq54}).

\section{Casimir force in D+1 dimensions}

Our consideration so far have applied to $1$ and $3$ space
dimensions. For some purposes it is, however, useful to
generalize the preset calculations to the $D+1$ space-time
dimensions. To avoid any ambiguity it is easier to develop the
abstract concept of $D$ dimensional formalism by comparing it
with the $3$ dimensional formulation. The boundaries are now
nonpenetrating hyperplates of thickness $d$ whose interior and
exterior hyperplanes of interfaces are located at $x_1=\pm a/2$
and $x_1=\pm (a/2+d)$, respectively. We label the hyperplates as
well as the different regions of this geometry, by analogy with
Fig.~\ref{figure}, by $1$ and $2$ as well as $1$, $2$ and $3$, respectively.

It is understood that the field operator in the region $3$ is now
in the form of

\begin{equation}
\label{eq82}{\hat \varphi}\left( {\bf r},t \right)=
\left(\frac{\hbar c^2}{2}\right)^{1/2}\int d^{(D-1)} {\bf
p}_{\parallel}\int_{0}^{\infty} \frac{d p_1}{\sqrt {\omega}_p
}{\hat a}\left({\bf p }, t\right)u \left({\bf p},{\bf r} \right)
+\mathrm{H.c.}
\end{equation}
where ${\bf r}$ and ${\bf p}$ are $D$-vectors in $D$ dimensional
coordinate and momentum space. The mode function $u \left({\bf
p},{\bf r} \right)$ is in the form of

\begin{equation}
\label{eq83}u\left({\bf p, {\bf r}}\right)=(2 \pi
\hbar)^{(1-D)/2}\exp(i{\bf p}_{\parallel}\cdot {\bf
x}_{\parallel}/\hbar)u(p_1, x_1)
\end{equation}
where ${\bf x}_{\parallel}=(x_2, x_3,...x_D)$, ${\bf
p}_{\parallel}=(p_2, p_3,...p_D)$ and $u(p_1, x_1)$ is given by
Eq.~(\ref{eq12}). Likewise, we can write down the field operator in the region $2$ with the help of Eqs.~(\ref{eq40}) and (\ref{eq41}) along with the
obvious changes which are related the dimensional considerations.
Therefore

\begin{equation}
\label{eq84}{\hat \varphi}\left({\bf r
},t\right)=\left(\frac{\hbar c^2}{2}\right)^{1/2}\int d^{(D-1)}
{\bf p}_{\parallel} \sum_{\Omega, n}\frac{1}{\sqrt
{{\omega}_n^{\Omega} }}{\hat a}_n^{\Omega}\left({\bf
p}_{\parallel}, t\right)u_n^{\Omega} \left({\bf p}_{\parallel},
{\bf r} \right)+H.c.
\end{equation}
where

\begin{equation}
\label{eq85}u_n^{\Omega} \left({\bf p}_{\parallel}, {\bf r}
\right)=\frac{1}{(2 \pi \hbar)^{(D-1)/2}}\exp(i{\bf
p}_{\parallel}\cdot {\bf x}_{\parallel}/\hbar)u_n ^{\Omega}(x_1)
\end{equation}
Note that $u_n ^{\Omega}(x_1)$ is given by Eqs.~(\ref{eq13}) and (\ref{eq14}).

The field operators show that the calculations are very similar
to the $3$ dimensional formalism. The net force on the hyperplate
$2$ can, therefore, be evaluated in the usual manner and
expressed as

\begin{equation}
\label{eq86}F=\frac{1}{(2 \pi \hbar)^{(D-1)}} \int d^{(D-1)} {\bf
p}_{\parallel} \left\{ \frac{ \pi^2 \hbar^2 c}{2a^3}\left[\sum_{
n=0}^{\infty}f(n, p_{\parallel})-\int_{0}^{+\infty }dn f(n,
p_{\parallel})\right]\right\}
\end{equation}
where $f(n)$ is given by Eq.~(\ref{eq50}). Comparing Eq.~(\ref{eq49}) with Eq.
(\ref{eq86}), it is seen that the latter expression is obtained by the
well-known replacement

\begin{equation}
\label{eq87} \frac{1}{(2 \pi \hbar)^{2}}\int d^2 {\bf
p}_{\parallel}\rightarrow\frac{1}{(2 \pi \hbar)^{(D-1)}}\int
d^{(D-1)} {\bf p}_{\parallel}
\end{equation}
We can now use the poisson's sum formula along with the integral
representation of the modified Bessel function $K_0$ to write the
net force in the form of

\begin{equation}
\label{eq88}F=\frac{1}{(2 \pi \hbar)^{(D-1)}} \int d^{(D-1)} {\bf
p}_{\parallel} \left\{ -\frac{\hbar c}{4 \pi a^2}\sum_{
n=1}^{\infty}\frac{\partial^2}{\partial n^2}K_0
\left(\frac{2na}{\hbar}\sqrt{p_{\parallel}^2+m^2
c^2}\right)\right\}
\end{equation}
The integration on the hyperplane ${\bf p}_{\parallel}$ is all
that need be done. Before evaluating the integral, it is
instructive to treat the problem by the alternative approach
based on the fluctuation dissipation theorem and Kubo's formula.

By analogy with the $3$ dimensional formalism, it is seen that
the symmetry of the present configuration allows us to express
the coordinate space Green function in terms of its Fourier
transform as

\begin{equation}
\label{eq89} G({\bf r}, {\bf r}^{\prime}, \omega_p)=\frac{1}{(2
\pi \hbar)^{(D-1)}} \int d^{(D-1)} {\bf p}_{\parallel} G({\bf
p}_{\parallel}, x_1, x_1^{\prime}, \omega_p) \exp[i{\bf
p}_{\parallel}\cdot ({\bf x}_{\parallel}-{\bf
x}_{\parallel}^{\prime})/ \hbar]
\end{equation}
where $G({\bf p}_{\parallel}, x_1, x_1^{\prime}, \omega_p)$ is
given by Eqs.~(\ref{eq59}) and (\ref{eq60}) for the regions $3$ and $2$,
respectively. Note that this is in agreement with Eq.~(\ref{eq87}). In
general, the calculation of the Fourier transform integral over
the $D-1$ dimensional hyperplane ${\bf p}_{\parallel}$ will be
much more difficult than the calculation of the same integral over
the plane ${\bf p}_{\parallel}$. The reason for greater
difficulty is apparently that, in general, the more number of
dimensions involved the more difficulty to perform the
integration. Fortunately, it is the imaginary part of the space
Green function at ${\bf r}={\bf r}^{\prime}$, not the explicit
form of it, that concern us. Therefore, recalling the discussion
above Eq.~(\ref{eq78}), it is seen that regardless of how complicated the
space Green functions are in the region $3$ and $2$, the net
force on the hyperplate $2$ is obtained by substitution of Eq.~(\ref{eq89}) into Eq.~(\ref{eq79}). We stressed that for the first term of Eq.~(\ref{eq79}), the Fourier transformed Green function is given by Eq.
(\ref{eq60}), while for the second term it is given by Eq.~(\ref{eq59}). By
combining Eqs.~(\ref{eq89}) and (\ref{eq79}) and making use of ${\bf
x}_{\parallel}={\bf x}_{\parallel}^{\prime}$ on both of the
interior and exterior hyperplanes of the hyperplate $2$, we find
easily that

\begin{eqnarray}
\label{eq90}
&&F=\frac{1}{(2 \pi \hbar)^{(D-1)}}\int d^{(D-1)} {\bf
p}_{\parallel}\frac{\hbar}{2
\pi}\int_{mc^2/\hbar}^{\infty}d\omega_p \left\{\left[
\frac{\partial}{\partial x_1}\frac{\partial}{\partial
x_1^{\prime}} {\mathrm{Im}}G\left({\bf r}, {\bf r
}^{\prime},\omega_p\right) \right ]_{x_1=x_1^{\prime}=a/2}\right.\nonumber\\
&&\hspace{3in}\left.-\left[
\frac{\partial}{\partial x_1}\frac{\partial}{\partial
x_1^{\prime}} {\mathrm{Im}}G\left({\bf r}, {\bf
r}^{\prime},\omega_p\right) \right
]_{x_1=x_1^{\prime}=a/2+d}\right\}.
\end{eqnarray}
Inserting Eqs.~(\ref{eq60}) and (\ref{eq59}) into Eq.~(\ref{eq90}), and taking into
account that the bulk part contributions in the first and second
terms of Eq.~(\ref{eq90}) cancel each other. Furthermore, on expanding
the prefactor of the square brackets in Eq.~(\ref{eq60}) and by comparing
with the $1$ and $3$ dimensional formalism, it is seen that the
reflection term in Eq.~(\ref{eq59}) and the first term in Eq.~(\ref{eq60}) with
$n=0$ cancel each other as well. The result can, therefore, be
written down as

\begin{equation}
\label{eq91}F=\frac{1}{(2 \pi \hbar)^{(D-1)}}\int d^{(D-1)} {\bf
p}_{\parallel}\left(\frac{c}{ \pi \hbar}\right)
\sum_{n=1}^{\infty}\int_{0}^{\infty}dp_1 \frac{p_1^2
\cos\left(\frac {2np_1 a}{\hbar}\right)}{\sqrt{p_1^2
+P_{\parallel}^2 + m^2 c^2}}
\end{equation}
In writing Eq.~(\ref{eq91}) we apply a suitable change of variable of
integration, that is $p_1$ instead of ${\omega}_p$. Employing the
integral representation of the modified Bessel function $K_0$,
Eq.~(\ref{eq91}) can be written in the form of Eq.~(\ref{eq88}). It is,
therefore, seen that the result obtained in this way fully agrees
with result that can be obtained using the method of field
quantization.

The integration in Eq.~(\ref{eq88}) over the hyperplane ${\bf
p}_{\parallel}$ can be done easily in the usual manner. The method
rests on the definition of the solid angle in $D-1$ dimensional
space of the hyperplane ${\bf p}_{\parallel}$. Since the integrand
possesses spherical symmetry in the hyperplane, it depends only
on $p_{\parallel}$, the integration over the solid angle in the
momentum space ${\bf p}_{\parallel}$ is trivial. We find that

\begin{equation}
\label{eq92}F=-\frac{ \hbar c}{4 \pi a^2}\frac{1}{(2 \pi
\hbar)^{(D-1)}}\frac{(\sqrt {\pi})^{(D-1)}}{\Gamma (\frac
{D-1}{2})}\sum_{n=1}^{\infty} \frac{\partial^2}{\partial n^2 }\int
dp_{\parallel}{p_{\parallel}}^{(D-2)}K_0
\left(\frac{2na}{\hbar}\sqrt{p_{\parallel}^2+m^2 c^2}\right)
\end{equation}
where $\Gamma$ is the Gamma function. There remains only the
integration on $p_{\parallel}$. Despite the complicated
dependence of the integrand on $p_{\parallel}$, the integral is
well-known and yields

\begin{equation}
\label{eq93}F=-\frac{ \hbar c}{4 \pi a^2}\left(\frac{\xi}{2a \sqrt
{2 \pi} }\right)^{(D-1)}\sum_{n=1}^{\infty}
\frac{\partial^2}{\partial n^2 }\left(\frac{1}{n
\xi}\right)^{(D-1)/2}K_{(D-1)/2} (n \xi)
\end{equation}
Using the recurrence relations of the modified Bessel functions,
we can easily show that

\begin{equation}
\label{eq94}F=-2 \hbar c\left(\frac{\xi}{2a \sqrt {2 \pi}
}\right)^{(D+1)}\sum_{n=1}^{\infty} \left(\frac{1}{n
\xi}\right)^{(D+1)/2}\left[(n \xi)K_{(D+3)/2} (n \xi)-K_{(D+1)/2}
(n \xi)\right]
\end{equation}
We can now derive the Casimir force for a neutral massless scalar
field. This is obtained by inserting the limiting forms of the
modified Bessel functions in Eq.~(\ref{eq94}). The result can be
expressed as

\begin{equation}
\label{eq95}F=-\frac{D \Gamma[(D+1)/2] 
\zeta(D+1) \hbar c}{(2a
\sqrt { \pi})^{(D+1)}}
\end{equation}
where $\zeta$ is the Zeta function. It is evident that by setting
$D=1$ and $D=3$ in Eq.~(\ref{eq95}), one can easily obtain Eqs.~(\ref{eq34}) and
 (\ref{eq55}), respectively. The latter equation is a well known result
which has been reported in the literature \cite{ref6}.
\section{Conclusion}

We have presented two approaches to the calculation of the
Casimir force for massive scalar field. The more formal and
illustrative one relies on the general tool of the field theory,
i.e. the explicit form of the quantum field operators, while the
other way, perhaps concise and more simple, originates from the
fluctuation dissipation theorem and Kubo's formula. Apparently,
the latter provides an alternative formulation of the effect that
mutually equivalent.

To simplify the calculation, we consider the attractive force
between two nonpenetrating plates. The scope of the presented
paper is extended to the derivation of the formal expression of
the force in three different spatial dimensions. We begin with
1+1 space-time dimensions whose force expression is given in 
Eq.~(\ref{eq33}). The details of the calculations are written in an
appropriate form which can easily be extended to 3+1 space-time
dimensions whose force expression is provided in Eq.~(\ref{eq54}). It is
seen that the latter calculation is the straight forward
generalization of the former. This allows one to recast this
formalism into $D+1$ space-time dimensions in a similar manner
whose force expression is of the form Eq.~(\ref{eq94}).

If we restrict our discussion to massless scalar field, we are
thus led to cast the main result of the present work into a brief
and concise statement that the force between two nonpenetrating
plates is always in the form of $C_{D} \left( \hbar c/
a^{D+1}\right)$, where $a$ is the separation between the two
plates, $D$ denotes to the spatial dimensions of the problem and
the constant factor $C_{D}$ is all that is needed to be obtained
by the details of the calculations. This can also be verified by
using the dimensional considerations.

\acknowledgements
I would like to thank R.~Matloob and H.~Safari for very helpful comments and suggestions without which the present work would not have been refined.
{}



\end{document}